\documentclass[aps,twocolumn,secnumarabic,superscriptaddress,amssymb,amsmath,nobibnotes,aps,prl,floatfix]{revtex4-1}
\usepackage{docs}%
\usepackage{bm}%
\usepackage[scr,scaled=1.1]{rsfso}
\usepackage[export]{adjustbox}
\usepackage[colorlinks=true,linkcolor=blue]{hyperref}%
\usepackage{graphicx}
\usepackage{dutchcal} 
\expandafter\ifx\csname package@font\endcsname\relax\else
 \expandafter\expandafter
 \expandafter\usepackage
 \expandafter\expandafter
 \expandafter{\csname package@font\endcsname}%
\fi

\def\e{\,{\rm e}}

\def\kk#1#2{k_{#1}\cdot k_{#2}}

\def\half{{1\over 2}}

\def\Z{{\mathchoice {\hbox{$\sf\textstyle Z\kern-0.4em Z$}}
{\hbox{$\sf\textstyle Z\kern-0.4em Z$}}
{\hbox{$\sf\scriptstyle Z\kern-0.3em Z$}}
{\hbox{$\sf\scriptscriptstyle Z\kern-0.2em Z$}}}}

        \def\slash#1{#1\!\!\!\raise.15ex\hbox {/}}
\newcommand{\slD}{\,\raise.15ex\hbox{$/$}\kern-.27em\hbox{$\!\!\!D$}}
\newcommand{\slpartial}{\raise.15ex\hbox{$/$}\kern-.57em\hbox{$\partial$}}

\def\epsk#1#2{\varepsilon_{#1}\cdot k_{#2}}
\def\epseps#1#2{\varepsilon_{#1}\cdot\varepsilon_{#2}}
\def\Gd{\dot{G}}

\def\no{\noindent}

\def\be{\begin{equation}}
\def\ee{\end{equation}\noindent}
\def\bear{\begin{eqnarray}}
\def\ear{\end{eqnarray}\noindent}
\def\bec{\blue\begin{equation}}
\def\eec{\end{equation}\black\noindent}
\def\bearc{\blue\begin{eqnarray}}
\def\earc{\end{eqnarray}\black\noindent}
\def\benn{\begin{enumerate}}
\def\enn{\end{enumerate}}
\def\ee{&=&}

\def\tr{{\rm tr}\,}

\def\e{\,{\rm e}}

\def\b0{{\bf 0}}
\def\at{\left.\vphantom{\int}\right|}

\def\4piTD{{(4\pi T)}^{-{D\over 2}}}
\def\4piT4{{(4\pi T)}^{-2}}

\begin{document}

\title{Color-kinematics duality from the Bern-Kosower formalism}

\author{Naser Ahmadiniaz}
\email{n.ahmadiniaz@hzdr.de}
\affiliation{Helmholtz-Zentrum Dresden-Rossendorf, Bautzner Landstra\ss e 400, 01328 Dresden, Germany}
\author{Filippo Maria Balli}
\email{filippo.balli@unimore.it}
\affiliation{Dipartimento di Scienze Fisiche, Informatiche e Matematiche, Universit\`{a} degli Studi di Modena e Reggio Emilia, Via Campi 213/A, I-41125 Modena, Italy\\
and INFN, Sezione di Bologna, Via Irnerio 46, I-40126 Bologna, Italy}
\author{Cristhiam Lopez-Arcos}
\email[Corresponding author:\,]{cmlopeza@unal.edu.co}
\affiliation{Escuela de Matem\'{a}ticas, Universidad Nacional de Colombia Sede Medell\'{i}n, Carrera 65 $\#$ 59A--110, Medell\'{i}n, Colombia}
\author{Alexander Quintero V\'{e}lez}
\email{aquinte2@unal.edu.co}
\affiliation{Escuela de Matem\'{a}ticas, Universidad Nacional de Colombia Sede Medell\'{i}n, Carrera 65 $\#$ 59A--110, Medell\'{i}n, Colombia}
\author{Christian Schubert}
\email{schubert@ifm.umich.mx}
\affiliation{Instituto de F\'{i}sica y Matem\'{a}ticas Universidad Michoacana de San Nicol\'{a}s de Hidalgo
Edificio C-3, Apdo. Postal 2-82 C.P. 58040, Morelia, Michoac\'{a}n, M\'{e}xico}

\begin{abstract}
Berends-Giele currents are fundamental building blocks for on-shell amplitudes in non-abelian gauge theory.
We present a novel procedure to construct them using the Bern-Kosower formalism for one-loop gluon amplitudes. Applying
the pinch procedure of that formalism to a suitable special case the currents are naturally obtained in terms of 
multi-particle fields and obeying colour-kinematics duality. As a feedback to the Bern-Kosower formalism we outline how the multi-particle polarisations and field-strength tensors 
can be used to significantly streamline the pinch procedure.

\end{abstract}

\maketitle

\section{Introduction}\label{sec:intro}
In non-abelian gauge theory, the nonlinearity of the Yang-Mills gauge transformations tends to make it difficult to write $n$-gluon amplitudes in a way that would make the gauge Ward identities transparent. 
In the off-shell case, these identities are inhomogeneous, relating the $n$-gluon amplitudes to the $(n-1)$-gluon ones 
(see \cite{Ahmadiniaz:2020htz} and refs. therein). 
On-shell they imply transversality of the corresponding $n$-gluon matrix elements, but in the standard Feynman diagram approach this involves intricate cancellations between one-particle irreducible diagrams and reducible diagrams that have trees sewn onto loops.  Such trees, with any number of gluons on-shell and one off-shell (sewn onto the loop)  are called Berends-Giele currents. Berends and Giele already introduced the idea of interpreting them in terms of multi-gluon states \cite{Berends:1987me}. More recently, it was found that these currents can be written in a certain gauge, the so-called BCJ gauge \cite{Lee:2015upy}, motivated by a result of Bern, Carrasco and Johansson that allows one to write the numerators for the color ordered amplitudes in such a way that these obey the same relations as their color factors \cite{Bern:2008qj}. This has come to be known as color-kinematics duality. 
At the level of the Berends-Giele currents, the duality is made explicit by means of some generalized Jacobi identities (GJI), as has been shown by Mafra and Schlotterer \cite{Mafra:2014oia}. They arrived at these identities by carefully looking at the structure of multi-particle vertex operators for ten dimensional Super Yang-Mills (SYM), making use of the technology of pure spinor BRST cohomology. 
 
The study of the Berends-Giele currents has recently received a good deal of attention \cite{Mafra:2015gia,Mafra:2015vca,Lee:2015upy,Berg:2016wux,Berg:2016fui,Garozzo:2018uzj,Bridges:2019siz,Edison:2020uzf}, since writing amplitudes in this BCJ form is the key to exhibiting colour-kinematics as well as double-copy duality \cite{Kawai:1985xq,Bern:2010ue}. However, their construction  quickly becomes tedious in practice as the number of legs grows, and sophisticated methods have been proposed to deal with this problem (see, e.g., \cite{Lee:2015upy,Bridges:2019siz} for an approach involving highly non-linear gauge transformations resembling pure spinor BRST transformations).

In the present work, we will use the Bern-Kosower formalism to develop a simple and direct method to construct the currents in the BCJ gauge. 
This formalism was originally derived using the field-theory limit of string amplitudes \cite{Bern1991-1669,bern1991color,bern1992computation}, and led to a set of rules that allows one to directly write down Feynman-Schwinger type parameter integrals for the one-loop on-shell $n$-gluon matrix elements. For our present purposes, the most relevant aspect of these Bern-Kosower rules is that they allow one to reconstruct the integrands of the reducible contributions 
to the matrix elements from the one of the irreducible one by a pinching procedure. The application of the Bern-Kosower pinch rules requires one to first perform certain partial integrations to the integrand that effectively remove quartic vertices. 
The general structure of the resulting integrands was studied by Strassler \cite{Strassler:1992zr,Strassler:1992nc}
in the framework of the worldline formalism, an alternative approach to perturbation theory that to some extent mimics string perturbation theory (for reviews see \cite{Schubert2001-73,Edwards:2019eby}). He found that the partial integration procedure naturally leads to the appearance of ``Lorentz cycles'' $Z_k(i_1,i_2,\ldots ,i_k)$, defined as traces of products of gluon field strength tensors
$f_i^{\mu\nu} \equiv k_i^\mu\varepsilon_i^\nu - \varepsilon_i^\mu k_i^\nu$. 
%
This fact has already turned out to be extremely useful for the derivation of form-factor decompositions of the three- and four-gluon vertex functions \cite{Ahmadiniaz:2012xp,Ahmadiniaz:2016qwn}, and even in the abelian case leads to more compact integral representations for
 photonic processes than had been previously known \cite{Schmidt:1994aq,Ahmadiniaz:2020jgo}.

Clearly, the pinching procedure must hold the full information on the Berends-Giele currents attached to the loop. Our present work is motivated by
the observation that it leads to the appearance of generalized structures involving now also multi-particle fields. We observe that the pinching procedure naturally generates these multi-particle fields in the BCJ gauge. 
We explicitly calculate the currents up to the five-point case, finding them to obey the required GJI.


\paragraph{Notation:}
For a word $I=i_1 i_2\dots i_p$ we set $\vert I\vert=p$. We also have multiparticle momentum $k_{I}^{\mu}=k_{i_1}^{\mu}+k_{i_2}^{\mu}+\dots+k_{i_p}^{\mu}$ and Mandelstam variables $s_I=k_I^2$. 
We write Lorentz cycles involving multi-particle fields numerators as
\bear\label{eq:worded-LC}
Z_k(I_1,\dots,I_k) &=& \Bigl(\frac{1}{2}\Bigr)^{\delta_{k2}}\mathrm{tr}\bigg(\prod_{i=1}^{k}f_{I_i}\bigg).
\ear


\section{Symmetric partial integration and cycle decomposition}

Central to the Bern-Kosower formalism is the following master formula for the
color-ordered one-loop $n$-gluon correlator with a scalar loop, 
\begin{widetext}
\begin{eqnarray}
\Gamma(k_1,\varepsilon_1;\ldots;k_n,\varepsilon_n)
&=&
{(-ig)}^n
\tr (T^{a_1}\cdots T^{a_n})
{\int_{0}^{\infty}}dT
{(4\pi T)}^{-\frac{D}{2}}
\e^{-m^2T}
\int_0^T d\tau_1 \int_0^{\tau_1}d\tau_2 \cdots \int_0^{\tau_{n-2}} d\tau_{n-1}
\nonumber\\ &&\times 
\exp \biggl\lbrace \sum_{i,j=1}^n 
\Bigl(  \half G_{ij} k_i\cdot k_j
-i\dot G_{ij}\varepsilon_i\cdot k_j
+\half\ddot G_{ij}\varepsilon_i\cdot\varepsilon_j
\Bigr)\biggr\rbrace
\Bigl\vert_{\varepsilon_1\ldots \varepsilon_n}
\label{master}
\end{eqnarray}
\end{widetext}
\no
Here $G_{ij}\equiv G(\tau_i,\tau_j)$ are the worldline Green's function 
$G(\tau,\tau') = \vert \tau -\tau'\vert -\frac{(\tau-\tau')^2}{T}$, with an antisymmetric derivative ($\Gd_{ij}=-\Gd_{ji}$). 
For a given $n$, we expand the exponential keeping only the terms
linear in each of the polarization vectors $\varepsilon_1,\ldots,\varepsilon_n$. 
The resulting integrand is of the form 
\bear \exp\biggl\lbrace 
\cdot
\biggr\rbrace 
\Bigl\vert_{\varepsilon_1\varepsilon_2\ldots \varepsilon_n}
 \equiv
 {(-i)}^n P_n(\dot G_{ij},\ddot G_{ij})
 {\rm e}^{\half \sum_{i,j=1}^n G_{ij}k_i\cdot k_j }
\nonumber\\
 \label{defPN}
 \ear\no
 with polynomials $P_n$.

\begin{figure}[h]
\centering 
\includegraphics[scale=0.85]{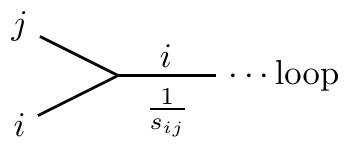}
\caption{\label{fig-pinch} Pinching of a vertex according to the Bern-Kosower rules.}
\end{figure}

As it stands, \eqref{master} represents the off-shell irreducible vertex function induced by a scalar loop.
To obtain the corresponding
on-shell matrix elements, one normally would have to perform a Legendre transformation, which diagrammatically
means attaching trees to the loop in all possible ways.
Alternatively, in the Bern-Kosower formalism these one-particle reducible contributions can be included 
by the following pinching procedure:
(i) Remove all second derivatives $\ddot G_{ij}$ contained in $P_n$ through integration by parts. 
(ii) Draw all possible $\phi^3$ 1-loop diagrams $D_i$ with $n$ legs, labelled $1,\ldots,n$ (following the
ordering of the color trace). 
(iii) A diagram will contribute iff each vertex except the ones attached to the loop corresponds to a possible pinch.
A vertex with labels $i<j$ can be pinched iff $P_n$ is linear in $\dot G_{ij}$. The pinching replaces this $\dot G_{ij}$ by
a factor of $1/s_{ij}$, removes the vertex and transfers the label $i$ to the ingoing leg (see Fig. \ref{fig-pinch}).  

The $\tau_j$ - integration is omitted and the index $j$ replaced by $i$ in all $G_{kl},\dot G_{kl}$. 
The trees are to be pruned recursively starting with the outermost vertices. 
For details and examples see \cite{bern1992computation,Schubert2001-73}.

The result of the integration-by-parts procedure is ambiguous starting from the four-gluon case \cite{Strassler:1992nc}.
For our present purposes, it is imperative to fix this ambiguity in a way that preserves the full permutation symmetry of the
master formula \eqref{master}. A suitable ``symmetric partial integration'' algorithm was proposed in \cite{Schubert:1997ph}. 
It transforms the polynomial $P_n(\dot G_{ij},\ddot G_{ij})$ into a polynomial $Q_n(\dot G_{ij})$ that permits a
cycle decomposition into terms with a definite cycle content, each cycle corresponding to a product
\bear
\dot G(i_1,i_2,\cdots ,i_k) = \dot G_{i_1i_2} 
\dot G_{i_2i_3} 
\cdots
\dot G_{i_ki_1}
Z_k(i_1,i_2,\cdots ,i_k) \nonumber\\
\label{defbicycle}
\ear
with $Z_k(i_1,i_2,\cdots ,i_k)$ like in \eqref{eq:worded-LC}.  
These cycles are invariant under the operations of cyclic permutations and
inversion. 
Apart from cycle factors a term may have a left-over, called \emph{tail} and denoted by $T(i_1,i_2,\cdots ,i_k)$
where $i_1,i_2,\ldots,i_k$ are the indices not yet bound up in any cycle. 
Although the tails are not manifestly transversal, they turn into total derivatives whenever any of the $\varepsilon_{i_m}$
contained in them is replaced by $k_{i_m}$. 
For example, the cycle decompositions of $Q_3$ and $Q_4$ read
\begin{align}\label{Q3}
\begin{split}
Q_3&=Q_3^3+Q_3^2\\
Q_3^3 &=
\dot G(1,2,3)
\\
Q_3^2 &=
\dot G(1,2)T(3)+\dot G(2,3)T(1)+\dot G(3,1)T(2)
\end{split}
\end{align}
\bear
Q_4&=&Q_4^4+Q_4^3+Q_4^2+Q_4^{22}\nonumber\\
Q_4^4 &=& 
\dot G(1,2,3,4)+\dot G(1,2,4,3) + \dot G(1,3,2,4)
\nonumber\\
Q_4^3 &=&
\dot G(1,2,3)T(4)+\dot G(2,3,4)T(1)+\dot G(3,4,1)T(2)\nonumber\\
&&+\dot G(4,1,2)T(3)
\nonumber\\
Q_4^2 &=&
\dot G(1,2)T(3,4)+\dot G(1,3)T(2,4)+\dot G(1,4)T(2,3)
\nonumber\\ &&
+\dot G(2,3)T(1,4)+\dot G(2,4)T(1,3)+\dot G(3,4)T(1,2)
\nonumber\\
Q_4^{22} &=&
\dot G(1,2)\dot G(3,4)+\dot G(1,3)\dot G(2,4)+\dot G(1,4)\dot G(2,3)
\nonumber\\
\label{Q4}
\ear
where the cycle content of a term is indicated by the superscript. 
At this level, only the one- and two tails appear,
\begin{align}\label{def2tail}
\begin{split}
&T(a) \equiv \sum_r\dot G_{ar} \varepsilon_a\cdot k_r  
\\
&T(a,b) \equiv
\sum_{{r,s}\atop {(r,s)\ne (b,a)}}
\dot G_{ar}
\varepsilon_a\cdot k_r
\dot G_{bs}
\varepsilon_b\cdot k_s
\\ 
&\:\:+\half
\dot G_{ab}
\varepsilon_a\cdot\varepsilon_b
\Bigl[
\sum_{r\ne b}
\dot G_{ar}k_a\cdot k_r 
- \sum_{s\ne a} \dot G_{bs}k_b\cdot k_s
\Bigr]
\end{split}
\end{align}
In the two-tail, note the exclusion of terms from the sums that would correspond to the appearance of a new cycle $\dot G(a,b)$ in the tail, and thus to an overcounting. Note also that when advancing from the $n$-gluon amplitude to the $n+1$-gluon one the only new
ingredient to be calculated is the $n-1$ tail, a relatively easy task. 
The integrands $Q_5$ and $Q_6$ are explicitly shown in Appendix C of \cite{Schubert2001-73}.


\section{The structure of worldline integrands and pinch operators}\label{sec:BKprime-rules}
The main ingredients for our approach are the Bern-Kosower rules \cite{Bern1991-1669,bern1992computation,bern1991color} and the symmetric partial integration \cite{Schubert:1997ph}. From these we identify two essential objects. First, we have the permutation invariant integrand $Q_n$. Second, for two adjacent legs $i$ and $j$ with $i < j$, an antisymmetric \emph{pinch operator} acting on $Q_n$ as
\begin{equation}\label{eq:brack-op}
\mathscr{D}_{ij}Q_n=\frac{\partial}{\partial\Gd_{ij}}Q_n\at_{\substack{\Gd_{ij}=0 \phantom{iiii}  \\ \Gd_{jk}\rightarrow\Gd_{ik}}}.
\end{equation}
Diagrammatically, this corresponds to pinching the two adjacent legs $i$ and $j$, see Fig.~\ref{fig-pinch}. 

In order to understand the basic link between the former and the GJI, let us first define what they are. Consider the free Lie algebra $\mathrm{Lie}[1,\dots,n]$ generated by all words in the letters in $1,\dots, n$ (see, e.g., \cite{GARSIA1990309,reutenauer1993free}). Moreover, let $\ell$ be the left-to-right bracketing on $\mathrm{Lie}[1,\dots,n]$, which is defined recursively by
\begin{align*}
\begin{split}
\ell(i_1 i_2 \cdots i_k) &= \ell(i_1 i_2 \cdots i_{k-1})i_k - i_k \ell(i_1 i_2 \cdots i_{k-1}), \\
\ell(i) &= i, \\
\ell(\varnothing) &= 0.
\end{split}
\end{align*}
Then the GJI of order $k$ can be characterized as the set of identities in $\mathrm{Lie}[1,\dots,n]$ of the form
\begin{equation}\label{eq:GJI}
I \ell(J) + J \ell(I) = 0,
\end{equation}
for every pair of non-empty words $I$ and $J$ such that $\vert I \vert +  \vert J \vert =k$.

Now that we have the GJI, we can explore the connection between these and the pinch operators in more detail. For this, we need to understand a little better the structure of $Q_n$. Given a bijection or two-to-one map $\alpha \colon \{1,\dots,n\} \to \{1,\dots,n\}$, consider the following polynomial of degree $n$ on the $\Gd_{ij}$'s
\begin{equation}\label{eq:poly}
Q_n^{(\alpha)} = \sum_{\mathrm{perm.}} c_{12\dots n} \Gd_{1 \alpha(1)} \Gd_{2 \alpha(2)} \cdots \Gd_{n \alpha(n)},
\end{equation}
where the coefficients depend on the polarizations and momenta. Then, $Q_n$ can be written as a sum of polynomials of this form. Therefore, to understand how the pinch operators act on $Q_n$, it will be enough to consider their action on such polynomials. For this, it will be convenient first to examine an specific example. Take $n = 4$ and $\alpha$ such that $\alpha(1) = 1$, $\alpha(2) = 1$, $\alpha(3) = 2$ and $\alpha(4) = 3$. By a straightforward calculation one finds that the action of the pinch operator $\mathscr{D}_{12}$ on the resulting polynomial $Q_{4}^{(\alpha)}$ yields the polynomial
\begin{align}\label{eq:D12poly4}
\begin{split}
 & (c_{3142} - c_{3241}) \Gd_{13}^2 \Gd_{14} + (c_{4312} - c_{4321}) \Gd_{13} \Gd_{34}^2  \\
 & + (3\leftrightarrow 4).
\end{split}
\end{align} 
We can check directly that each of the coefficients is antisymmetric in $1$ and $2$, i.e., they satisfy the GJI of order $1$. Let us next apply the pinch operator $\mathscr{D}_{13}$ to \eqref{eq:D12poly4}. The result is
\begin{equation}
(c_{4132} - c_{4231} + c_{4312} - c_{4321}) \Gd_{14}^2.
\end{equation}
Now the coefficient satisfies the Jacobi identity in $1$, $2$ and  $3$, i.e., the GJI of order $2$. 

Returning to the general case, we may infer that the iterated action of the pinch operators $\mathscr{D}_{12},\mathscr{D}_{13},\dots, \mathscr{D}_{1(n-1)}$ on a polynomial of the form \eqref{eq:poly} will produce a monomial. Explicitly,
\begin{equation}\label{eq:maxpinch}
\mathscr{D}_{1(n-1)}\cdots \mathscr{D}_{13}\mathscr{D}_{12}Q_n^{(\alpha)}=\tilde{c}_{12\cdots n}\Gd_{1n}^2,
\end{equation}   
where the coefficient $\tilde{c}_{12\cdots n}$ satisfies the GJI of order $n-1$ in $1, 2,\dots, n-1$. We have checked that this is the case up to degree $n=8$.

We make a final remark on how the above can be represented diagrammatically. To this end, one just have to remark that a left-to-right bracketing in $\mathrm{Lie}[1,\dots,n]$ can be interpreted as a planar binary tree  and vice versa. For example, 
\begin{align*}
\ell(12) = \includegraphics[scale=0.5,valign=c]{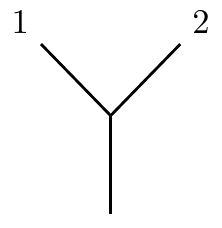}, &\quad \ell(123) = \includegraphics[scale=0.5,valign=c]{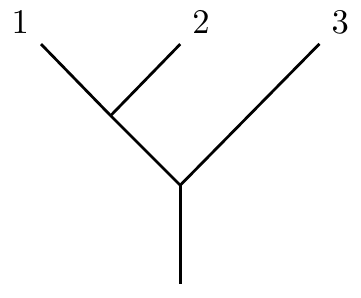}, \quad \text{etc.}
\end{align*}
Using this notation, we find, for instance, that the iterated action of $\mathscr{D}_{1,2}$ and $\mathscr{D}_{1,3}$ on $Q_{n}$ can be graphically represented as
\begin{equation}\label{eq:D23action}
\mathscr{D}_{13}\mathscr{D}_{12}\left(\includegraphics[scale=0.45,valign=c]{{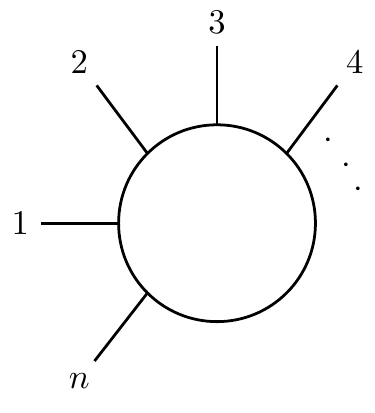}} \right) = \includegraphics[scale=0.45,valign=c]{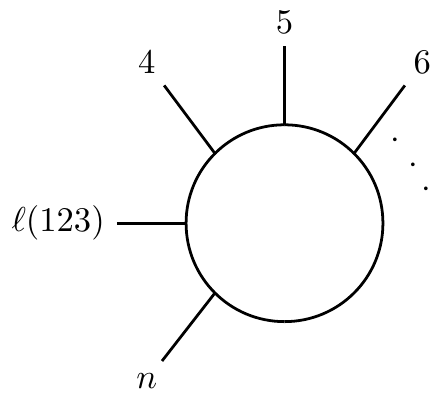}.
\end{equation}


\section{Multi-particle fields from pinching}

The properties of the Bern-Kosower integrands
discussed in the previous section allow us to build the Berends-Giele currents for YM directly in the BCJ gauge. The tree propagators could be obtained by the conventional Bern-Kosower pinch rules, but we prefer to exploit the symmetries in the numerators that allow one to obtain all the numerators of the Berends-Giele currents from a single calculation, by relabeling the legs and taking linear combinations where necessary. For the propagators we can make use of the inverse of the KLT matrix \cite{Kawai:1985xq,Bern:1998sv,BjerrumBohr:2010hn}, which was given a Berends-Giele description in \cite{Mafra:2016ltu,Mafra:2020qst} (see also \cite{Frost:2020eoa}).

The appropriate polynomial for finding the field strength numerator is just the sum of the terms in $Q_n$ with a single one-cycle component $\tilde{Q}_n = Q_n^2+Q_n^3+ \ldots + Q_n^n$. Using \eqref{eq:maxpinch} we obtain   
\bear\label{eq:BGF-natural}
\mathscr{D}_{1(n-1)}\cdots \mathscr{D}_{13}\mathscr{D}_{12}\tilde{Q}_n &=& Z_2(12\dots n-1,n)\Gd_{1n}^2\nonumber\\
&=&\tfrac{1}{2}f_{12\cdots (n-1)}^{\mu\nu}f_{n\nu\mu}\Gd_{1n}^2
\ear
where $f_{12\cdots (n-1)}^{\mu\nu}$ satisfy the GJI. 

From the field-strength tensors one can also extract the multi-particle polarizations, but it turns out that those can alternatively 
be obtained applying pinch operators just to the tails
\bear\label{eq:BGE-natural}
\mathscr{D}_{1(n-2)}\cdots \mathscr{D}_{13}\mathscr{D}_{12}T(1,2,\dots,n-2)&=&\nonumber\\
&&\hspace{-4cm}\varepsilon_{12\cdots (n-2)}\cdot k_{n-1}\Gd_{1(n-1)} + \varepsilon_{12\cdots (n-2)}\cdot k_{n}\Gd_{1n}.\nonumber\\
\ear  
The multi-particle polarizations obtained in either way will satisfy the corresponding GJI. Note, however, that the second \emph{tail-pinching} method requires one to know the tails to one order higher than is necessary for the first \emph{cycle-pinching} approach.


\section{Examples}\label{sec:examples}

We will now work out the currents up to $n=5$ (the five-point case is in the Appendix). 
We start with $Q_2$ which has no trees and therefore no derivatives, but leads to the first solution
\bear
Z_2(1,2)\Gd_{12}^2.
\label{two-point}
\ear
Comparing (\ref{two-point}) with \eqref{eq:worded-LC} one arrives at 
\bear
Z_2(1,2)=\tfrac{1}{2}f_{1}^{\mu\nu}f_{2\nu\mu},
\ear
which gives us just the usual abelian one-particle field strength tensor
$f_i^{\mu\nu}=k_i^{\mu}\varepsilon_i^{\nu} - k_i^{\nu}\varepsilon_i^{\mu}$
that was already introduced above. This is also the Berends-Giele current for this case.

\subsection{Two-particle case}

This numerator is extracted from the $Q_3$ integrand \eqref{Q3}. To obtain the external-leg bubble integrand numerator we only have to pinch two legs
\bear
Q_3^{(12)}	= \mathscr{D}_{12}Q_3.
\ear
The explicit expression for the Lorentz two-cycle that goes with $\Gd_{13}^2$ in \eqref{eq:BGF-natural} for this case is
\bear
Z_2(12,3)=\epsk21 Z_2(1,3) - \tfrac{1}{2}Z_3(1,2,3) - (1\leftrightarrow 2)\,.\nonumber\\ 
\ear
We can immediately see that $f_3^{\nu\mu}$ can be factorized out to give the two-current field strength  numerator
\bear\label{eq:f12}
f_{12}^{\mu\nu}=\epsk21 f_1^{\mu\nu} - (f_1f_2)^{\mu\nu} - (1\leftrightarrow 2)\, .
\ear
From \eqref{def2tail} and \eqref{eq:BGE-natural} we extract the two-particle polarization 
\bear\label{eq:e12}
\varepsilon_{12}^{\mu}= \tfrac{1}{2}\left[\epsk21\varepsilon_1^{\mu} - \varepsilon_{1\rho}f_{2}^{\rho\mu} - (1\leftrightarrow 2)\right].
\ear
Evidently $f_{12}^{\mu\nu}$ and $\varepsilon_{12}^{\mu}$ are antisymmetric in 1 and 2.
The Berends-Giele current here is 
\bear\label{eq:BG-2}
\mathscr{F}^{\mu\nu}_{12} = \frac{f^{\mu\nu}_{12}}{s_{12}},
\ear
which can also be obtained directly from the classical action via perturbiner methods \cite{Mafra:2016ltu,Mizera:2018jbh,Lopez-Arcos:2019hvg}. Such methods will not work at higher points if we look for BCJ gauge. 

\subsection{Three-particle case}

Here $\mathscr{D}_{13}\mathscr{D}_{12}\tilde{Q}_4$ will immediately drop the numerator for the field strength three-current	
\bear\label{eq:f123}
f_{123}^{\mu\nu} &=& k_{123}^{\mu}\varepsilon_{123}^{\nu} -k_{12}\cdot k_3\varepsilon_{12}^{\mu}\varepsilon_{3}^{\nu}\nonumber\\ 
&&- k_{1}\cdot k_2(\varepsilon_{1}^{\mu}\varepsilon_{23}^{\nu}+\varepsilon_{13}^{\mu}\varepsilon_{2}^{\nu})-(\mu\leftrightarrow\nu).
\ear

For the polarization we use the expression for $T(1,2,3)$ that can be found in the Appendix C of \cite{Schubert2001-73} and by \eqref{eq:BGE-natural} 
\begin{eqnarray}\label{eq:eps123}
\varepsilon_{123}^\mu &=&\tfrac{1}{2}\big[\left(k_3\cdot\varepsilon_{12}\right)\varepsilon_3^\mu-\left(k_{12}\cdot \varepsilon_3\right)\varepsilon_{12}^\mu+\varepsilon_{12 \nu}\,f_{3}^{\nu\mu}-\varepsilon_{3\nu}\,f_{12}^{\nu\mu}\nonumber\\
&&+\tfrac{1}{2}\varepsilon_1\cdot \varepsilon_2\,\varepsilon_3\cdot \left(k_1-k_2\right)k_{123}^{\mu} \big]\,.
\end{eqnarray}
The term in the second line of (\ref{eq:eps123}) to be called $h_{123}$
\bear\label{eq:h3}
h_{123}=\tfrac{1}{4}\varepsilon_1\cdot \varepsilon_2\,\varepsilon_3\cdot \left(k_2-k_1\right),
\ear
is directly related to the transformation in \cite{Mafra:2014oia} that takes the numerators from the Lorenz gauge to the BCJ gauge, which appears naturally in our method. Both \eqref{eq:f123} and \eqref{eq:eps123} satisfy the following GJI
\bear\label{eq:lie3}
\varepsilon_{123}^{\mu}+\varepsilon_{213}^{\mu}=0\quad ,\quad \varepsilon_{123}^{\mu} + \varepsilon_{312}^{\mu} + \varepsilon_{231}^{\mu}=0.
\ear  
Now the Berends-Giele current for the polarization looks like
\bear
\mathscr{A}_{123}^{\mu}=\frac{\varepsilon_{123}^{\mu}}{s_{12}s_{123}} + \frac{\varepsilon_{321}^{\mu}}{s_{23}s_{123}}.
\ear

\subsection{Four-particle case}

Following the same procedure we arrive at the field strength numerator in the four-particle case
\bear\label{eq:f1234}
f_{1234}^{\mu\nu}&=& k_{1234}^{\mu}\varepsilon_{1234}^{\nu}+k_{123}\cdot k_4\varepsilon_{123}^{\nu}\varepsilon_{4}^{\mu}+k_{12}\cdot k_{3}\big[\varepsilon_{12}^{\nu}\varepsilon_{34}^{\mu}\nonumber\\
&&+\varepsilon_{124}^{\nu}\varepsilon_{3}^{\mu}\big]+k_{1}\cdot k_{2}\big[\varepsilon_{1}^{\nu}\varepsilon_{234}^{\mu}+\varepsilon_{134}^{\nu}\varepsilon_{2}^{\mu}+\varepsilon_{13}^{\nu}\varepsilon_{24}^{\mu}\nonumber\\
&&+\varepsilon_{14}^{\nu}\varepsilon_{23}^{\mu}\big]-(\mu\leftrightarrow\nu).
\ear
We also find the numerator of the four-particle polarization 
\begin{eqnarray}\label{eq:e1234}
\varepsilon_{1234}^\mu &=& \tfrac{1}{2}\big[\varepsilon_4^\mu\left(\varepsilon_{123}\cdot k_4\right)-\varepsilon_{123}^\mu\left(\varepsilon_{4}\cdot k_{123}\right)+\varepsilon_{123\nu}f_4^{\nu\mu}\nonumber\\
&&-\varepsilon_{4\nu}f_{123}^{\nu\mu}\big]-\varepsilon_3^{\mu}(k_{12}\cdot k_3)h_{124}- k_1\cdot k_2\big(\varepsilon_2^{\mu}h_{134}\nonumber\\
&&-\varepsilon_1^\mu h_{234}\big)-k_{1234}^\mu h_{1234},
\end{eqnarray}
where  
\begin{eqnarray}\label{eq:h4}
h_{1234}&=&\tfrac{1}{4}\big[\varepsilon_1\cdot\varepsilon_2\,\varepsilon_3\cdot k_2 \varepsilon_4\cdot \,\left(k_1-k_{23}\right) \\
&+&\tfrac{1}{2}\left(\varepsilon_1\cdot \varepsilon_2\,\varepsilon_3\cdot\varepsilon_4 \, k_2\cdot k_3   \right)-(123\rightarrow 312)\big] \,- (1\leftrightarrow 2).\nonumber
\end{eqnarray}
A numerator with the BCJ property was also obtained in \cite{Mafra:2014oia} with a two-step procedure involving a BRST-inspired transformation.
The GJI satisfied by these numerators are
\bear
&&f_{1234}^{\mu\nu}+f_{2134}^{\mu\nu}=0,\quad f_{1234}^{\mu\nu} + f_{3124}^{\mu\nu} + f_{2314}^{\mu\nu}=0,\nonumber\\
&&f_{1234}^{\mu\nu} - f_{1243}^{\mu\nu} + f_{3412}^{\mu\nu} - f_{3421}^{\mu\nu}=0.
\ear 
The corresponding current is given by
\bear\label{eq:BG-4a}
\mathscr{F}_{1234}^{\mu\nu}&=&\frac{1}{s_{1234}}\Bigg(\frac{f_{1234}^{\mu\nu}}{s_{12}s_{123}} + \frac{f_{3214}^{\mu\nu}}{s_{23}s_{123}} + \frac{f_{1234}^{\mu\nu}-f_{1243}^{\mu\nu}}{s_{12}s_{34}}\nonumber\\
&&- \frac{f_{4321}^{\mu\nu}}{s_{34}s_{234}} - \frac{f_{2341}^{\mu\nu}}{s_{23}s_{234}}\bigg).
\ear


\section{Building amplitudes with multi-particle cycles and tails}

The trees that we have found can be attached back to the propagator to compute tree-level amplitudes with color-kinematics duality,
or we can just use the current numerators to build local BCJ numerators for the color-ordered amplitudes (for this type of procedures in tree-level amplitudes and examples see e.g. \cite{Mafra:2015vca,Gomez:2020vat}). 

But let us now return to the one-loop gluon amplitudes. Our calculations above have taught us that the pinching of 
the ordinary Lorentz cycles produces smaller cycles with insertions of multi-particle field-strength tensors, and the pinching
of tails leads to lower-point tails involving multi-particle polarizations. Moreover, for the cycle part such insertions have
already shown to be useful and natural for various amplitudes in SYM theory: generalized two-cycles
already appeared in six-dimensional SYM in \cite{Berg:2016wux,Berg:2016fui}, there called ``scalar fundamentals",
and a multi-particle version of the $t_8$ tensor, which in $\mathscr{N}=4$ SYM absorbs the tensor structure of the four-gluon amplitude,
was introduced in \cite{Edison:2020uzf}. 

This leads us to conjecture that the complete effect of the pinching procedure in the Bern-Kosower formalism
may, in the $Q_n$ integrand, be taken into account simply by adding to the un-pinched integrand 
all combinations of generalized cycles and
tails that are possible for the given number of gluons, and compatible with the fixed color ordering. Thus we now define a new
generalized Lorentz cycle
\bear\label{eq:generalized-LC}
\mathcal{Z}_k(I_1,\dots,I_k) &\equiv& \Bigl(\tfrac{1}{2}\Bigr)^{\delta_{k2}}\mathrm{tr}\bigg(\prod_{i=1}^{k}\mathscr{F}_{I_i}\bigg)\,,
\ear
which, contrary to \eqref{eq:worded-LC}, now uses the full Berends-Giele currents $\mathscr{F}_{I_i}$, and the
generalized tail
\bear\label{eq:generalized-tail}
\mathcal{T}_k(I_1,\dots,I_k) &\equiv & T(k_{I_1},\mathscr{A}_{I_1};\dots;k_{I_k},\mathscr{A}_{I_k})
\ear
For example, the three-gluon amplitude, whose un-pinched integrand is \eqref{Q3},
should have the pinch contribution
\bear
\dot G(1,23) + \dot G(2,31) + \dot G(3,12)
\ear
where, e.g.,
\bear
\dot G(1,23) = \dot G_{12}\dot G_{21} \half \tr \bigl(f_1\mathscr{F}_{23}\bigr)
\ear
and this is indeed what the Bern-Kosower pinch rules produce. 
Similarly, at the four-point level the prediction for the single-pinch terms would be
\bear
\dot G(1,2,34) + \dot G(1,23)T(4) + \dot G(1,2){\cal T}(34)+{\rm perm.}
\nonumber\\
\ear
and for the double pinches,
\bear
\dot G(1,234) + \dot G(12,34) +{\rm perm.}
\ear
which we have again found to be in agreement with the application of the pinch rules. 

This result for the double pinches can be straightforwardly generalized to the maximal, i.e. (n-2)  - fold, 
pinch contribution to the $n$-gluon amplitude. This is because the result of such a pinch can only
be a bubble diagram, and the bubbles can always be represented as a Lorentz two-cycle times 
a $\dot G_{ij}^2$, as we saw in \eqref{eq:BGF-natural} and have used above. 
For the bubble integrands we then have the compact expression for a given ordering $I$, namely
\bear\label{eq:n-bubble}
B_n^I&=&\sum_{\substack{I=JK\\|I|,|J|\geq2}}\sum_{\sigma\in\mathrm{cyc}(I)}\mathcal{Z}_2(\sigma_{j_1}\cdots\sigma_{j_{n-l}},\sigma_{k_{n-l+1}}\cdots\sigma_{k_n})\nonumber\\
&&\hspace{2cm}\times\Gd_{\sigma_{j_1}\sigma_{k_{n-l+1}}}^2\,,
\ear
where the first sum accounts for the deconcatenation of the word $I$ into words $J$ and $K$ of length bigger than one.

\section{Summary and Outlook}

We have presented a novel method of constructing Berends-Giele currents using the Bern-Kosower formalism and
a specific pinch contribution to the $n$-gluon amplitudes. Explicit calculation up to the five-point case has shown that
the multi-particle fields numerators of these currents obey the GJI required by BCJ gauge, indicative of color-kinematics duality. We are currently working on a formal proof for the all-$n$ case, based on our understanding of the general structure of the polynomials presented here. 

A connection between the Bern-Kosower formalism and color-kinematics duality was previously hinted at in \cite{Casali:2020knc} 
where it was shown that all the Bern-Kosower numerators satisfy BCJ identities through a detailed analysis of the field-theory limit of the monodromy relations of string theory at one loop; it would be interesting to compare their results with ours.

We have further shown that using these Berends-Giele currents as words in generalized Lorentz cycles,
and the associated multi-particle polarization vectors in generalized tails,
provides an extremely attractive approach towards absorbing the effect of the 
Bern-Kosower pinching procedure into multi-particle tensor structures. 
We hope to obtain along these lines a representation of the one-loop $n$-gluon amplitudes 
that would be ultracompact as well as exhibit manifest color-kinematics duality. 

\clearpage \onecolumngrid
\appendix

\section{Five-Particle field strength current}

In this section we use the procedure to compute the multi-particle field strength tensor to provide the full expression of the five-particle current numerator which leads to

\bear
f_{12345}^{\mu\nu}&=& -\epsk{5}{1234}f_{1234}^{\mu\nu} - \epsk{4}{123}f_{1235}^{\mu\nu} - \epsk{3}{12}f_{1245}^{\mu\nu} - \epsk{5}{123}\epsk{4}{123}f_{123}^{\mu\nu} - \frac{1}{2}\epseps{4}{5}\epsk{4}{123}f_{123}^{\mu\nu}\nonumber\\
&&+ \left[\epsk{1}{2}f_{2345}^{\mu\nu} + \epsk{2}{1}\epsk{3}{1}f_{145}^{\mu\nu} + \epsk21\epsk31\epsk41\epsk51f_{1}^{\mu\nu} - (1\leftrightarrow 2)\right]\nonumber\\
&&+\bigg[\Big\lbrace\epsk32\epsk{5}{124}f_{214}^{\mu\nu} - \frac{1}{2}\epseps35\kk{2}{3}f_{124}^{\mu\nu} - \epsk32\epsk{4}{12}f_{125}^{\mu\nu} - \epsk32\epsk41\epsk{5}{1}f_{12}^{\mu\nu}\nonumber\\
&&-\epsk32\epsk41\epsk{5}{2}f_{12}^{\mu\nu} -\frac{1}{2}\epseps35\epsk41\kk{2}{3}f_{12}^{\mu\nu} -\frac{1}{2}\epseps45\epsk32\kk{1}{4}f_{12}^{\mu\nu}\nonumber\\
&&-(123\rightarrow 312)\Big\rbrace - (1\leftrightarrow 2)\bigg] +\bigg[\Big\lbrace\epsk32\epsk42\epsk{5}{1}f_{21}^{\mu\nu} -\frac{1}{2}\epseps15\epsk32\epsk42\kk{1}{2}f_2^{\mu\nu}\nonumber\\
&&- (123\rightarrow 312)- (1234\rightarrow 4123)\Big\rbrace - (1\leftrightarrow 2)\bigg] +\bigg[\frac{1}{2}\Big\lbrace\epseps34\kk{2}{3}f_{215}^{\mu\nu}- \epseps34\epsk51\kk{2}{3}f_{12}^{\mu\nu} \nonumber\\
&&-\epseps34\epsk12\epsk52\kk{2}{3}f_{2}^{\mu\nu} -\frac{1}{2}\epseps15\epseps34\kk12\kk{2}{3}f_{2}^{\mu\nu} - (123\rightarrow 312)\nonumber\\
&&- (1234\rightarrow 4123)+ (1234\rightarrow 4312)\Big\rbrace - (1\leftrightarrow 2)\bigg]+\bigg[\Big\lbrace\epsk32\epsk42\epsk{5}{2}f_{21}^{\mu\nu} \nonumber\\
&&-\frac{1}{2}\epseps35\epsk42\kk{2}{3}f_{12}^{\mu\nu}-\frac{1}{2}\epseps45\epsk32\kk{2}{4}f_{12}^{\mu\nu} +\frac{1}{8}\epseps24\epseps35\kk12\kk{2}{3}f_1^{\mu\nu} \nonumber\\
&&+\frac{1}{8}\epseps23\epseps45\kk12\kk{2}{4}f_1^{\mu\nu}- (123\rightarrow 312)- (1234\rightarrow 4123)- (12345\rightarrow 51234)\Big\rbrace\nonumber\\
&&- (1\leftrightarrow 2)\bigg]+\bigg[\Big\lbrace\frac{1}{2}\epseps34\epsk52\kk{2}{3}f_{21}^{\mu\nu} +\frac{1}{8}\epseps25\epseps34\kk12\kk{2}{3}f_1^{\mu\nu} - (123\rightarrow 312)\nonumber\\
&&- (1234\rightarrow 4123)+ (1234\rightarrow 4312)- (12345\rightarrow 51234) +(12345\rightarrow 53124)\Big\rbrace \nonumber\\
&&- (1\leftrightarrow 2)\bigg] +\bigg[\Big\lbrace\frac{1}{2}\epseps45\epsk34\kk{2}{3}f_{21}^{\mu\nu} +\frac{1}{2}\epseps45\epsk23\epsk34\kk{1}{2}f_1^{\mu\nu} \nonumber\\
&&-\frac{1}{2}\epseps45\epsk12\epsk34\kk{2}{3}f_2^{\mu\nu}-\frac{1}{8}\epseps23\epseps45\kk12\kk{3}{4}f_1^{\mu\nu} -(f_1f_2f_3f_4f_5)^{\mu\nu}\nonumber\\
&&- (123\rightarrow 312)- (1234\rightarrow 4123)+ (1234\rightarrow 4312)- (12345\rightarrow 51234) \nonumber\\
&&+(12345\rightarrow 54123) -(12345\rightarrow 54312) +(12345\rightarrow 53124)\Big\rbrace - (1\leftrightarrow 2)\bigg]\nonumber\\
\ear
It can be checked that the five-particle case also satisfies the GJI, i. e. 
\bear
&&f_{12345}^{\mu\nu}+f_{21345}^{\mu\nu}=0,\quad f_{12345}^{\mu\nu} + f_{31245}^{\mu\nu} + f_{23145}^{\mu\nu}=0,\nonumber\\
&&f_{12345}^{\mu\nu} - f_{12435}^{\mu\nu} + f_{34125}^{\mu\nu} - f_{34215}^{\mu\nu}=0,\nonumber\\
&&f_{12345}^{\mu\nu} - f_{12354}^{\mu\nu} + f_{45123}^{\mu\nu} - f_{45213}^{\mu\nu} - f_{45312}^{\mu\nu} + f_{45321}^{\mu\nu}=0.\nonumber\\
\ear

\bibliography{REFERENCES}

\end{document}